\documentclass{article}
\usepackage{spconf,amsmath,graphicx}
\usepackage{multirow}
\usepackage{booktabs,tabularx}
\usepackage{graphicx}
\usepackage{enumitem}
\usepackage{float}
\usepackage[vietnamese]{babel}
\usepackage{comment}
\usepackage{makecell}

\title{Improving Whispered Speech Recognition Performance using Pseudo-whispered based Data Augmentation}
\name{Zhaofeng Lin, Tanvina Patel, Odette Scharenborg}
\address{Multimedia Computing Group, Delft University of Technology, the Netherlands}
\copyrightnotice{978-1-6654-7189-3/22/\$31.00~\copyright2023 IEEE}

\begin{document}
%
\maketitle
\begin{abstract} 
Whispering is a distinct form of speech known for its soft, breathy, and hushed characteristics, often used for private communication. The acoustic characteristics of whispered speech differ substantially from normally phonated speech and the scarcity of adequate training data leads to low automatic speech recognition (ASR) performance. To address the data scarcity issue, we use a signal processing-based technique that transforms the spectral characteristics of normal speech to those of pseudo-whispered speech. We augment an End-to-End ASR with pseudo-whispered speech and achieve an 18.2\% relative reduction in word error rate for whispered speech compared to the baseline. Results for the individual speaker groups in the wTIMIT database show the best results for US English. Further investigation showed that the lack of glottal information in whispered speech has the largest impact on whispered speech ASR performance. 

\end{abstract}
\begin{keywords}
Whispered speech, pseudo-whisper, end-to-end speech recognition, wTIMIT, signal processing
\end{keywords}
\section{Introduction}
\label{sec:intro}
Whispered speech differs substantially from normally phonated speech. In whispered speech, vocal fold vibrations are absent, resulting in voicelessness \cite{tartter1989s}, which leads to reduced energy and intensity \cite{ITO2005139}. Additionally, the airflow in whispered speech is often greater than in normal speech, which leads to breathy speech \cite{konnai2017whisper}. 
Whispering is often used in private conversations or in libraries or meetings. Whispered speech also occurs in pathological speech contexts: Speech from individuals who face vocal system challenges such as diseases affecting the vocal folds or post-larynx surgery \cite{solomon1989laryngeal,cao2016recognizing} is often whisper-like. Moreover, whispering has been found to be beneficial in reducing or avoiding stuttering \cite{ingham2009measurement}. 

While human listeners can largely comprehend the linguistic information in whispered speech \cite{osfar2011articulation}, automatic speech recognition (ASR) systems face significant challenges in accurately transcribing whispered speech compared to normal speech \cite{ITO2005139}. The poorer performance of ASR systems on whispered speech is due to the large acoustic differences between whispered and normal speech, e.g., pitch differences due to reduced vocal fold vibrations \cite{lim2011computational}, up-shifted formant frequencies \cite{7178927, SHARIFZADEH2012e49}, wider formant bandwidth \cite{7178927}, reduced phonetic cues, and a lower signal-to-noise ratio \cite{ITO2005139}, and most importantly the limited amount of training data, which makes it challenging to develop robust ASR models tailored to whispered speech. 

Several approaches have been proposed in the literature to improve whispered speech ASR. For hybrid ASR systems, these include using Teager Energy Cepstral Coefficients instead of traditional Mel-frequency cepstral coefficients (MFCC) \cite{TASLP_TEcc_AE_2017}, showing large improvements for Serbian whispered speech. In \cite{7178927} the authors proposed a method to generate pseudo-whispered speech segments using denoising autoencoders showing considerable performance improvements on their own dataset. For End-to-End (E2E) systems, Chang \textit{et al.} \cite{chang9383595} showed that a system trained with a frequency-weighted SpecAugment, a frequency-divided Convolutional Neural Network extractor, a layer-wise transfer learning approach, and pre-training outperformed their baseline with about 44\% relative improvement in character error rate on whispered speech from wTIMIT \cite{lim2011computational}. The work in \cite{gudepu2020whisper} generated whispered speech from normal speech using Generative Adversarial Networks-based voice conversion (VC) techniques for training data augmentation obtaining the current best results on wTIMIT: 29.4\% word error rate (WER). Other methods used multimodal data including articulatory cues from motion data \cite{1415287,cao16_slpat} and visual information \cite{petridis2018visual}. Although these techniques show that it is possible to improve the recognition of whispered speech, there is still a performance gap with respect to normal speech. 

In this paper, we aim to
1) understand the (detrimental) effects of the specific acoustic characteristics of whispered speech on whispered speech ASR performance
2) and deal with the data scarcity problem by generating artificial whispered speech to augment the training data for improved E2E whispered speech ASR. To that end, we propose a handcrafted signal-processing method to convert normal speech to pseudo-whispered speech in two independent steps. These steps allow us to create ``intermediate forms'' of normal-to-whispered speech, which in turn allow us to investigate the effect of the specific acoustic characteristics of whispered speech on whispered speech ASR.

For our experiments, we use the whispered TIMIT speech dataset \cite{lim2011computational}, which has two speaker groups with different accents, North American English and Singaporean English. \cite{lim2011computational} showed that the whispered speech recognition performance was dependent on the accent group, with worse results for the Singaporean English speech. Here we, to the best of our knowledge, for the first time, investigate the effect of different accents (speaker groups) in E2E whispered speech ASR.

\section{Methodology}
Section 2.1 describes the three datasets that were used for testing, training, and for generating the pseudo-whispered speech in the experiments.
Section 2.2 provides a brief comparison of normal versus whispered speech. Section 2.3 describes our approach to the conversion from normal to pseudo-whispered speech. Section 2.4 explains the experimental setup.

\label{sec:methodology}

\subsection{Datasets}
\label{sec:datasets} 
\subsubsection{wTIMIT}
The Whispered TIMIT (wTIMIT) corpus \cite{lim2011computational} consists of 450 phonetically balanced sentences in both normal (wTIMIT-n) and whispered (wTIMIT-w) speech from speakers from two accent groups: US and Singaporean English with 28 (12 male and 16 female) and 20 (12 male and 8 female) speakers, respectively. wTIMIT thus consists of four speaker groups: $\mathbf{N_{US}}$: Normal speech with US accent; $\mathbf{N_{SG}}$: Normal speech with SG accent; $\mathbf{W_{US}}$: Whispered speech with US accent; $\mathbf{W_{SG}}$: Whispered speech with SG accent.

wTIMIT was originally partitioned into a training and test set \cite{lim2011computational}. To prevent overfitting our E2E models, a re-partitioning of wTIMIT into training, development, and test sets is needed.
Preliminary experiments performed in \cite{chang9383595} showed that a partitioning of the training and test data where there was no speaker overlap in the training and test set, degraded performance by approximately 10\% relatively compared to a partitioning of the training and test data where the same speaker could occur in both. This relatively small difference in performance was attributed to the pitch being mostly absent in whispered speech.
Also given the fact that partitioning by speakers can lead to less data that can be used as training data, \cite{chang9383595} suggested that prohibiting speaker overlap between the training and test sets is unnecessary. Following \cite{chang9383595}, we re-partitioned wTIMIT into a training, development, and test set allowing speaker overlap. Each data set consisted of 400/25/25 sentences, respectively, split from the 450 sentences.

\subsubsection{TIMIT}
The 450 prompts of wTIMIT were obtained from the phonetically balanced section (SX) of the TIMIT corpus \cite{garofolo1993darpa}, which makes TIMIT a good option as additional training data for normal speech and to generate pseudo-whispered speech. TIMIT contains US English read speech.
To validate the training process, we also evaluate our models on TIMIT test set.

\subsubsection{LibriSpeech}
The LibriSpeech corpus \cite{Libri7178964} consists of English read speech from audiobooks. LibriSpeech has recently been used as an additional (pre)training dataset in the recent E2E whispered speech studies \cite{chang9383595,gudepu2020whisper}. We therefore used its 100 hours subset to generate pseudo-whispered speech and augment the training data. 

\subsection{Differences between normal and whispered speech}
\label{sec:differenceNorvsWhisper}
\begin{figure}[htb]
\centering
\centerline{\includegraphics[width=0.9\columnwidth]{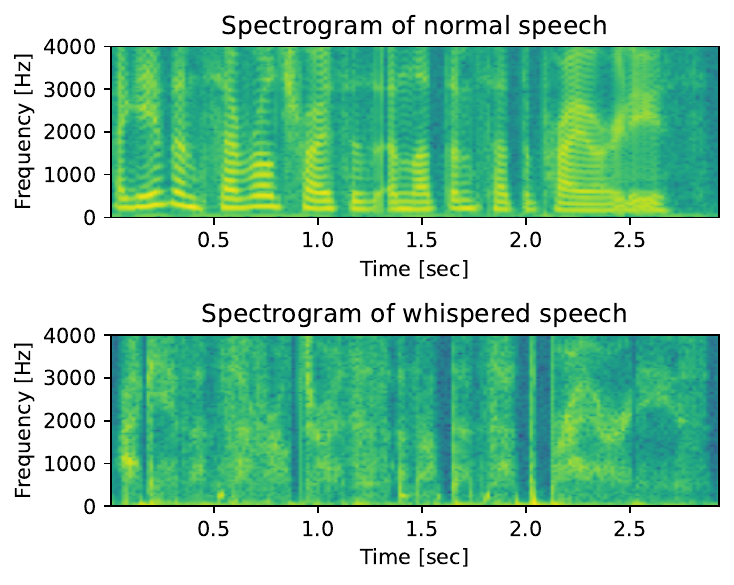}}
\caption{Spectrograms of the sentence ``I gave them several choices and let them set the priorities.'' produced by the same speaker in a normal and whispered voice. Example is taken from the wTIMIT corpus.}
\label{fig:sp}
\end{figure}

Figure \ref{fig:sp} shows the spectrograms of a female speaker from the wTIMIT corpus speaking the same utterance in a normal voice (top panel) and while whispering (bottom panel). Comparison of the spectrograms shows that whispered speech appears to have no formant information, less energy in particularly the lower frequency bands, and altogether a different acoustic profile compared to normal speech (see also \cite{chang9383595,tecc9287634,feng2022parameterization}). These differences are primarily due to the altered vocal production mechanism and reduced vocal fold activity during whispering \cite{lim2011computational}. Since in whispered speech voicing is significantly reduced or even absent, whispered speech clearly has less distinct harmonic patterns, which makes it challenging to identify individual formants in the spectrogram.

\subsection{Proposed pseudo-whispered speech conversion}

\begin{figure*}[ht]
\centering
\centerline{\includegraphics[width=0.8\textwidth]{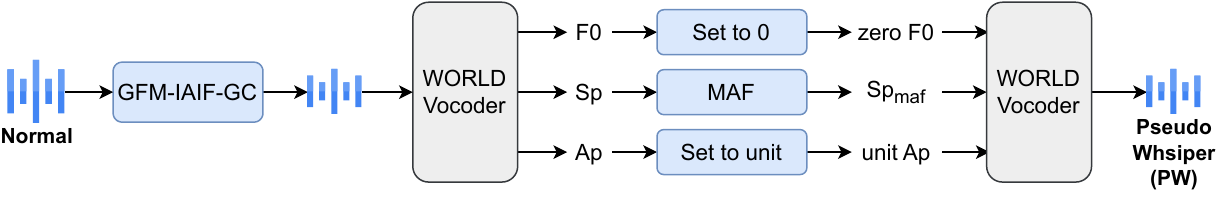}}
\caption{The proposed pipeline for pseudo-whispered speech conversion, where GFM-IAIF-GC is GFM-IAIF-based glottal cancellation and MAF is moving average filtering. Input is normal speech and the output is pseudo-whispered speech (PW).}
\label{fig:pw}
\end{figure*}

Our proposed method to convert normal to pseudo-whispered speech is based on that of Cotescu \textit{et al.} \cite{cotescu2019voice} who proposed a handcrafted digital signal processing (DSP) recipe that converts normal speech into whispered speech in three steps by making acoustic modifications to the normal speech: 
1) remove the glottal contribution using spectral subtraction; 2) shift the first formant using frequency warping; 3) increase the formant bandwidth using moving average filtering. The WORLD vocoder \cite{morise2016world} is used to extract features for re-synthesizing high-quality speech. 

In step 1, instead of using spectral subtraction as in \cite{cotescu2019voice}, we implemented a glottal cancellation method, which does not require parameters for modelling glottal flow but removes the glottal information directly from a given normal speech signal. Moreover, preliminary experiments using the method from \cite{cotescu2019voice} showed that moving average filtering not only widens the formant bandwidth but also up-shifts the formant frequencies. 
Hence, our proposed method for pseudo-whispered speech conversion is implemented in 2 steps, which is also shown in Figure \ref{fig:pw}:
\begin{enumerate}[itemsep=2pt,topsep=4pt,parsep=1pt]
   \item Removing the glottal source using GFM-IAIF-based Glottal Inverse Filtering (see section \ref{sec:gfm-iaif-gc}),
   \item Increasing the formant bandwidth and up-shifting the formant frequencies using moving average filtering (see section \ref{sec:maf}).
\end{enumerate}

\subsubsection{Step 1: Removing glottal information}
\label{sec:gfm-iaif-gc}

Under the assumption of the source-filter model \cite{fant1971acoustic}, a speech signal is composed of an excitation $E$, vocal tract filter $V$, lip radiation filter $L$, and glottis component $G$, which can be written in the frequency domain as $S(f) = E(f) \cdot G(f) \cdot V(f) \cdot L(f) $.
Glottal Inverse Filtering (GIF) estimates the source of voiced speech, specifically the glottal volume velocity waveform. 
Iterative Adaptive Inverse Filtering (IAIF) \cite{alku1992glottal} is one of the most widely used algorithms for GIF. IAIF successively models the vocal tract filter $V(f)$, lip radiation $L(f)$, and glottis $G(f)$ using linear prediction (LP) analysis, then removes their effect by inverse filtering. After two iterations, it ultimately removes $V(f)$ and $L(f)$ to leave an estimate of the glottal flow $g(n)$, where n is the discrete-time index.

The Iterative Adaptive Inverse Filtering method based on a
Glottal Flow Model (GFM-IAIF) \cite{8682625} is an improved version of IAIF. It constrains glottal flow by a $3^{rd}$ order spectral model $G(z)=\left\{\left(1-a z^{-1}\right)\left(1-a^* z^{-1}\right)\left(1-b z^{-1}\right)\right\}^{-1}$. 
GFM-IAIF performs competitively for normal phonations \cite{mokhtari2018estimation}, which makes it suitable for our case: cancelling glottal contribution of normally phonated speech.
In our method, we employ GFM-IAIF to extract glottal flow, after which the effect of the glottis is cancelled by inverse filtering, and the output is a speech signal without glottal contribution. 

Subsequently, $F0$, the spectral envelope ($Sp$), and the aperiodic spectral envelope ($Ap$) are extracted from the speech signal without glottal information using the WORLD vocoder. To ensure that the pitch is removed entirely, we set the $F0$ to zero and $Ap$ values to all units. 

\subsubsection{Step 2: Changing formant information}
\label{sec:maf}
To increase the formant bandwidth and up-shift the formant frequencies, we employ moving average filtering on the spectral envelop $Sp$ extracted by the WORLD vocoder with a 400 Hz-wide triangular window across all frequency axes and get a new spectrogram $Sp_{maf}$. 

\begin{figure}[htb]
\centering
\centerline{\includegraphics[width=1\columnwidth]{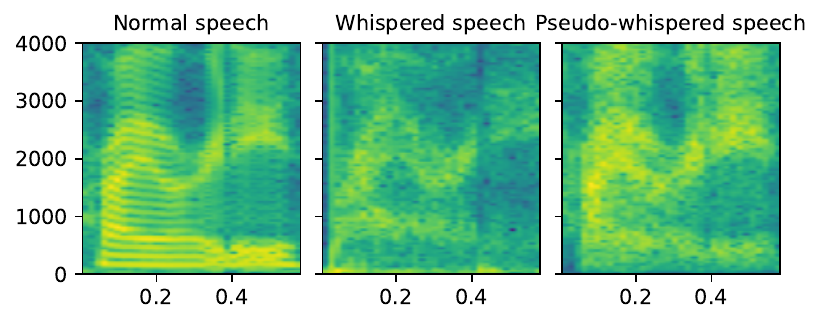}}
\caption{Spectrogram of normal (left panel), whispered (middle panel), and pseudo-whispered speech  (right panel) of the word ``priorities'' from the same utterance as in Figure 1.}
\label{fig:sp_3}
\end{figure}

The three adapted features, $zero \ F0$, $Sp_{maf}$ and $unit \ Ap$ are passed to the WORLD vocoder for re-synthesising the pseudo-whispered speech PW.
Figure \ref{fig:sp_3} shows an example of the conversion results, it shows a zoomed-in spectrogram of normal speech (left panel), whispered speech (middle panel) and pseudo-whispered speech (right panel). 

\subsection{Experimental setup}
\label{sec:exp_setup}

All E2E models were trained with the ESPNet toolkit \cite{watanabe2018espnet}. As speech files in TIMIT and LibriSpeech are recorded with a sampling rate of 16 kHz, all speech files in wTIMIT were downsampled from 44.1 kHz to 16 kHz. The front-end features are $80$ dimensional log-mel filterbank features with $3$-dimensional pitch features used for network training.

\subsubsection{Exp. 1: Baseline models}
First, we trained a strong baseline by investigating the
\begin{enumerate}[itemsep=2pt,topsep=4pt,parsep=1pt]
    \item Training data: TIMIT plus the normal speech from wTIMIT (TM+wTM-n) vs. TIMIT plus both normal and whispered speech from wTIMIT (TM+wTM-wn);

    \item Data augmentation: none vs. speed perturbation (SP) \cite{ko2015audio} at 90\% and 110\% of the original rate of the training data and SpecAugment \cite{park2019specaugment} which was used with a maximum width of each time and frequency mask of $T=20$, $F=10$, respectively;

    \item Dictionary types and sizes: six models used character-level dictionaries, and the remaining two used a Byte Pair Encoding (BPE) token dictionary \cite{sennrich2016neural};

    \item Model architectures: we compared the Hybrid-CTC \cite{watanabe2017hybrid} and Conformer \cite{guo2021recent} architectures (from the LibriSpeech recipe from the ESPNet framework). 
\end{enumerate}

In total, eight models were trained. The models were evaluated on the TIMIT and wTIMIT-n and wTIMIT-w datasets. Performance was measured in terms of Word Error Rate (WER) for both accent groups separately.

\subsubsection{Exp. 2: Pseudo-whisper data augmentation}
The second experiment investigated the effect of adding pseudo-whispered (PW) data to the training data on whispered speech ASR performance. To that end, the pseudo-whispered speech was created from TIMIT, wTIMIT-n, and LibriSpeech-100h and each was successively added to the training data: first only PW speech from TIMIT (PW(TM)), then the PW speech from wTIMIT-n (PW(wTM-n)) was added, and finally also the PW speech from LibriSpeech-100h (PW(Libri100)). Each set of training data was used to train both the Hybrid-CTC and Conformer architectures, yielding six models. The effect of the pseudo-whisper data augmentation on the two accent groups was also analysed.

\subsubsection{Exp. 3: Acoustic characteristics of whispered speech}
In the final experiment, the effect of the specific acoustic characteristics of whispered speech on whispered speech ASR performance was investigated by comparing the recognition results on speech in which either the glottal information was removed or in which the formant bandwidth had been widened and the formant frequencies shifted.

\begin{figure}[htb]
\centering
\centerline{\includegraphics[width=1\columnwidth]{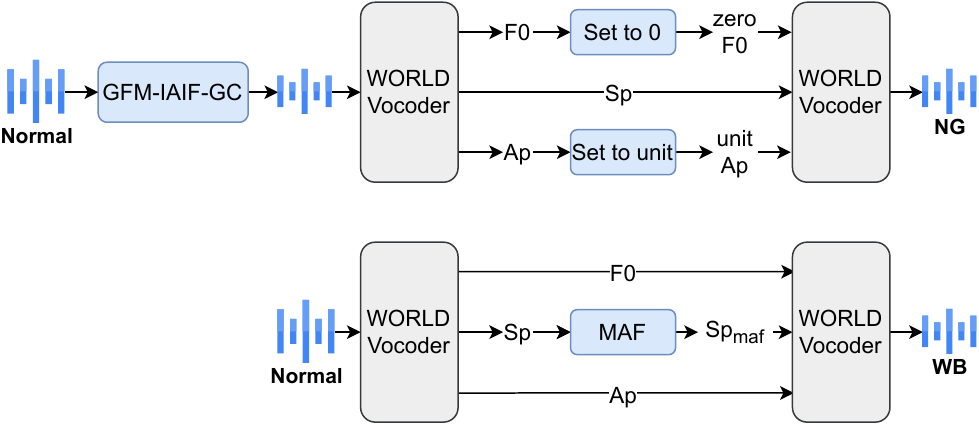}}
\caption{The pipeline for generating speech with only glottal cancellation (top panel) and with only a widened formant bandwidth and shifted formant frequencies (bottom panel).}
\label{fig:rq2}
\end{figure}

To that end, we individually applied each of the two steps of our proposed pseudo-whispered speech conversion method on the normal test set in wTIMIT and synthesized the modified speech. 
Figure \ref{fig:rq2} shows the pipelines of generating speech without glottal contribution (referred to as NG) and speech with widened formant bandwidth and shifted formant frequencies (referred to as WB). The pipelines are subsets of the full pipeline in Figure \ref{fig:pw}. 

The modified speech was subsequently tested using three models: the Hybrid-CTC architecture trained on only normal speech (row TM+wTM-n in Table \ref{tab:baseline}); trained on normal and whispered speech (row TM+wTM-wn in Table \ref{tab:baseline}); and trained on normal, whispered, and pseudo-whispered speech (TM+wTM-wn+PW(TM+wTM-n)). SP and SpecAug were not applied to all three models in this final experiment.

\begin{table*}[!ht]
\caption{WER (\%) of the eight baseline ASR systems on the TIMIT and wTIMIT test sets for the two accent groups separately. $\mathbf{N_{Avg}}$ is normal speech averaged; $\mathbf{W_{Avg}}$ is whispered speech averaged.} 
\label{tab:baseline}
\renewcommand{\arraystretch}{1}
\setlength{\tabcolsep}{0.6\tabcolsep} 
\resizebox{\linewidth}{!}{%
\begin{tabular}{@{}llcccc|c|cccc|cc@{}}
\toprule
\multicolumn{6}{c|}{\textbf{Details}} &
  \multicolumn{1}{c|}{\textbf{TM}} &
  \multicolumn{6}{c}{\textbf{wTIMIT}} \\ \midrule
\textbf{Training} &
  \textbf{Augmentation} &
  \multicolumn{1}{l}{\textbf{Hours}} &
  \multicolumn{1}{l}{\textbf{Architecture}} &
  \textbf{Token} &
  \multicolumn{1}{c|}{\textbf{\#Token}} &
  \multicolumn{1}{c|}{\textbf{Test}} &
  {$\mathbf{N_{US}}$} &
  {$\mathbf{N_{SG}}$} &
  {$\mathbf{W_{US}}$} &
  {$\mathbf{W_{SG}}$} &
  {$\mathbf{N_{Avg}}$} &
  {$\mathbf{W_{Avg}}$} \\ \midrule
\multirow{2}{*}{TM+wTM-n} &
  \multirow{2}{*}{None} &
  \multirow{2}{*}{28.9} &
  Hybrid-CTC &
  Char &
  29 &
  40.6 &
  45.4 &
  59.1 &
  99.4 &
  105.2 &
  51.2 &
  101.9 \\
 &
   &
   &
  Conformer &
  Char &
  29 &
  52.9 &
  73.4 &
  82.7 &
  102.5 &
  109.6 &
  77.4 &
  105.5 \\ \midrule
\multirow{2}{*}{TM+wTM-wn} &
  \multirow{2}{*}{None} &
  \multirow{2}{*}{55.1} &
  Hybrid-CTC &
  Char &
  29 &
  41.2 &
  51.5 &
  62.8 &
  55.3 &
  74.4 &
  56.3 &
  63.5 \\
 &
   &
   &
  Conformer &
  Char &
  29 &
  44.7 &
  78.0 &
  86.4 &
  81.4 &
  92.0 &
  81.6 &
  85.9 \\\midrule
\multirow{2}{*}{TM+wTM-wn} &
  \multirow{2}{*}{SP + SpecAug} &
  \multirow{2}{*}{166.3} &
  Hybrid-CTC &
  Char &
  29 &
  42.3 &
  52.0 &
  67.3 &
  57.0 &
  77.7 &
  58.5 &
  65.9 \\
 &
   &
   &
  Conformer &
  Char &
  29 &
  38.3 &
  49.6 &
  58.3 &
  53.2 &
  68.3 &
  53.3 &
  59.7 \\\midrule
\multirow{2}{*}{TM+wTM-wn} &
  \multirow{2}{*}{SP + SpecAug} &
  \multirow{2}{*}{166.3} &
  Hybrid-CTC &
  BPE &
  100 &
  44.6 &
  41.2 &
  55.8 &
  44.1 &
  65.9 &
  47.4 &
  53.4 \\
 &
   &
   &
  Conformer &
  BPE &
  100 &
  \textbf{34.1} &
  \textbf{34.9} &
  \textbf{41.8} &
  \textbf{37.7} &
  \textbf{53.5} &
  \textbf{37.9} &
  \textbf{44.4} \\ \bottomrule
\end{tabular}%
}
\end{table*}

\section{Results}
\label{sec:experiment}

\subsection{Exp. 1: Baseline models}
\label{sec:baseline}
Table \ref{tab:baseline} presents the results of the baseline models on the TIMIT and wTIMIT test sets for the two accent groups separately and averaged over both accent groups. Only training on normal speech (TM + wTM-n) gave a WER of 40\% (Hybrid-CTC) and 52\% (Conformer) on the TM test set, while the performance on wTIMIT-n averaged over both accent groups ($\mathbf{N_{Avg}}$) showed a fairly large WER drop of 10-20\%. The performance on whispered speech is much lower than that on normal speech, with WERs of over 100\%. 
Including whispered speech in the training data (TM+wTM-wn) improved recognition performance for whispered speech substantially, and reduced the gap with performance on normal speech to less than 10\% for both architectures, but at the cost of a slight increase in WER for normal speech. This clearly indicates that using matched training and test sets improves performance.  

Applying SP and SpecAug, surprisingly, did not yield improvements for the Hybrid-CTC model; however, it led to a substantial (more than 25\%) improvement for both normal and whispered wTIMIT speech for the Conformer model. This suggests that as the training data size increases, the Conformer models outperform Hybrid-CTC models. 

When using BPE, the Conformer outperformed the Hybrid-CTC on all three test sets overall and for the individual accent groups, achieving WERs of 37.9\% and 44.4\% for normal and whispered wTIMIT speech, respectively. These models were selected as our baseline models.

\subsection{Exp. 2: Pseudo-whisper data augmentation}
Table \ref{tab:pw} presents the results of the  pseudo-whispered speech experiments. For ease of comparison, the results of the baseline Hybrid-CTC and Conformer models are added (identical to those reported in Table \ref{tab:baseline}. Adding only 3 hours of pseudo-whispered data from TIMIT (with speed perturbation yielding 9 hours; see rows \textbf{PW(TM)}) improved the average WER of whispered speech compared to the baseline for both models, with the largest relative improvement for the Conformer model (18.2\%). 
Interestingly, adding pseudo-whispered speech also improved the WER on the normal wTIMIT speech was reduced for both models. 

Adding the pseudo-whispered speech of the wTIMIT-n training set (\textbf{PW(TM+wTM-n)}) further improved recognition performance for the Hybrid-CTC model but performance for the Conformer model deteriorated for the normal and whispered speech. Recognition performance on the TM test set was again similar to the baseline models. Further adding the PW speech from LibriSpeech (\textbf{PW(TM
+wTM-n+Libri100)}) gave the best recognition performance for normal wTIMIT speech, but it deteriorated the performance for the whispered speech. The best whispered speech results were obtained with the Conformer model trained with (only) the pseudo-whispered TIMIT speech added.

\begin{table*}[!ht]
\caption{WER (\%) on the TIMIT and wTIMIT test sets when using pseudo-whispered training data generated from TIMIT, wTIMIT-n, and LibriSpeech-100h. Relative improvement (\%) of the proposed method compared to the baseline is also reported.
Results of the chosen baseline Hybrid-CTC and Conformer models are added (identical to those reported in Table \ref{tab:baseline}.}
\label{tab:pw}
\setlength{\tabcolsep}{0.5\tabcolsep} 
\resizebox{\textwidth}{!}{%
\begin{tabular}{@{}lccc|c|cccccc|cc@{}}
\toprule
\multicolumn{4}{c|}{\textbf{Details}} &
  \textbf{TM} &
  \multicolumn{6}{c|}{\textbf{wTIMIT}} &
  \multicolumn{2}{c}{\textbf{Relative Imp.}} \\ \midrule
\textbf{Training Data} &
  \multicolumn{1}{l}{\textbf{Hours}} &
  \multicolumn{1}{l}{\textbf{Architecture}} &
  \textbf{\#Token} &
  \textbf{Test} &
  {$\mathbf{N_{US}}$} &
  {$\mathbf{N_{SG}}$} &
  {$\mathbf{W_{US}}$} &
  \multicolumn{1}{c|}{{$\mathbf{W_{SG}}$}} &
  {$\mathbf{N_{Avg}}$} &
  {$\mathbf{W_{Avg}}$} &
  {$\mathbf{N_{Avg}}$} &
  {$\mathbf{W_{Avg}}$} \\ \midrule
\multirow{2}{*}{\textbf{Baseline}} &
  \multirow{2}{*}{166.3} &
  Hybrid-CTC &
  100 &
  44.6 &
  41.2 &
  55.8 &
  44.1 &
  \multicolumn{1}{c|}{65.9} &
  47.4 &
  53.4 &
  - &
  - \\
 &
   &
  Conformer &
  100 &
  34.1 &
  34.9 &
  41.8 &
  37.7 &
  \multicolumn{1}{c|}{53.5} &
  37.9 &
  44.4 &
  - &
  - \\ \midrule
\multirow{2}{*}{+PW(TM)} &
  \multirow{2}{*}{175.8} &
  Hybrid-CTC &
  100 &
  46.5 &
  40.1 &
  52.4 &
  41.2 &
  \multicolumn{1}{c|}{59.9} &
  45.4 &
  49.2 &
  4.2 &
  7.9 \\
 &
   &
  Conformer &
  100 &
  36.4 &
  32.1 &
  \textbf{35.2} &
  33.4 &
  \multicolumn{1}{c|}{\textbf{40.1}} &
  33.4 &
  \textbf{36.3} &
  11.9 &
  \textbf{18.2} \\ \midrule
\multirow{2}{*}{+PW(TM+wTM-n)} &
  \multirow{2}{*}{253.6} &
  Hybrid-CTC &
  100 &
  43.4 &
  36.1 &
  44.3 &
  38.0 &
  \multicolumn{1}{c|}{51.8} &
  39.6 &
  43.9 &
  \textbf{16.5} &
  17.8 \\
 &
   &
  Conformer &
  100 &
  34.6 &
  32.4 &
  36.6 &
  33.6 &
  \multicolumn{1}{c|}{42.1} &
  34.2 &
  37.2 &
  9.8 &
  16.2 \\ \midrule
\multirow{2}{*}{+PW(TM+wTM-n+Libri100)} &
  \multirow{2}{*}{557.6

} &
  Hybrid-CTC &
  300 &
  16.9 &
  35.5 &
  55.0 &
  39.2 &
  \multicolumn{1}{c|}{63.4} &
  43.8 &
  \multicolumn{1}{c|}{49.5} &
  7.6 &
  7.3 \\
 &
   &
  Conformer &
  300 &
  \textbf{11.0} &
  \textbf{26.8} &
  38.6 &
  \textbf{30.7} &
  \multicolumn{1}{c|}{49.2} &
  \textbf{31.8} &
  \multicolumn{1}{c|}{38.6} &
  16.1 &
  13.1 \\ \bottomrule
\end{tabular}%
}
\end{table*}

\subsection{Analysis on different speaker groups}

Comparing the recognition performance on normal and whispered speech for the two accent groups in the wTIMIT test set showed that Singaporean English normal and whispered speech is consistently worse recognised than US English. This performance gap is the largest for whispered speech.

Adding pseudo-whispered speech always improved the recognition performance of normal and whispered US and Singaporean English, even if the pseudo-whispered speech was based on US English only (PW(TM)). In fact, adding only the US English pseudo-whispered speech from TM gave the best result for $\mathbf{N_{SG}}$ and $\mathbf{W_{SG}}$ and reduced the performance gap with US English to 3.1\% for normal speech and 7\% for whispered speech for the Conformer, i.e., the smallest performance gap for whispered speech for the Conformer.

Interestingly, adding pseudo-whispered speech from Singaporean English did not further improve recognition performance for $\mathbf{N_{SG}}$ and $\mathbf{W_{SG}}$ for the Conformer, although it did further improve performance for the Hybrid-CTC model, giving the best results for normal and whispered Singaporean English for the Hybrid-CTC model. 

When adding the pseudo-whispered US English from LibriSpeech (PW(Libri-100)) as additional training data, the performance on whispered US English ($\mathbf{W_{US}}$) improved to 30.7\%, the best result, but it adversely affected the performance on $\mathbf{W_{SG}}$, widening the gap between US and Singaporean English to almost 20\% for the Conformer model and even more for the Hybrid-CTC model. Thus, adding a large amount of pseudo-whispered speech based on US English negatively impacted the recognition of Singaporean English normal and whispered speech.

\subsection{Exp. 3: Acoustic characteristics of whispered speech}

Table \ref{tab:rq2} presents the results of the experiments on normal speech, real whispered speech, pseudo-whispered (PW) speech and the intermediate forms of whispered speech (see section \ref{sec:exp_setup}), i.e., normal speech without glottal contributions (NG) and normal speech with widened formant bandwidth and shifted formant frequencies (WB). Note that to create PW, both NG and WB are applied. 

When the model is trained on only normal speech, the gap between Normal and NG (>25\%) is larger than the one between Normal and WB (5\%). This indicates that performance is worse for speech without glottal contribution and that the widened formant bandwidth and shifted formant frequencies in whispered speech are less detrimental to recognition performance. Combining NG and WB into pseudo-whispered speech only shows a small deterioration compared to NG. This indicates that the effect of removing both glottal information and widening the formant bandwidth and shifting the formant frequencies is not entirely additive.

Not surprisingly, adding real whispered speech from wTIMIT-n greatly improves the recognition performance of real whispered speech. Recognition performance of pseudo-whispered speech and NG speech also greatly improves, to the level of that of real whispered speech. Performance on WB speech slightly deteriorates. This again indicates that the glottal information is the most important acoustic information to explain the whispered speech recognition performance. 

Adding pseudo-whispered speech improves recognition performance of real and pseudo-whispered speech, indicating that the pseudo-whispered speech is close enough to real whispered speech for real whispered speech to benefit from the added data. Recognition performance of PW is actually better than that of real whispered speech which shows the benefit of adding matched training data. Speech without glottal information is now actually better recognised than WB speech, which shows that adding speech without glottal information is most beneficial for NG speech and that the benefit for WB speech is less great. 

\begin{table}[h]
\centering
\caption{WERs (\%) of different test groups when the model is trained on normal speech (row TM+wTM-n in Table \ref{tab:baseline}); normal and whispered speech (row TM+wTM-wn in Table \ref{tab:baseline}); and normal, whispered, and pseudo-whispered speech (TM+wTM-wn+PW(TM+wTM-n)).}
\label{tab:rq2}
\setlength{\tabcolsep}{0.7\tabcolsep} 
\begin{tabular}{@{}lccccc@{}}
\toprule
\textbf{Training data} & \textbf{Normal} & \textbf{Whisper} & \textbf{PW} & \textbf{NG} & \textbf{WB} \\ \midrule
TM+wTM-n  & 51.2 & 101.9 & 79.7 & 78.0 & 56.5 \\

\midrule
TM+wTM-wn & 56.3 & 63.5  & 65.5 & 65.2 & 59.2 \\

\midrule
\begin{tabular}[c]{@{}l@{}}TM+wTM-wn\\ +PW(TM+wTM-n)\end{tabular} & 55.9            & 61.3            & 59.2        & 59.4        & 62.3        \\ \bottomrule
\end{tabular}
\end{table}

\section{Discussion and Conclusion}
\label{sec:conclusion}

This paper aims to deal with the data scarcity problem of whispered speech by generating artificial whispered speech to augment the training data for improved E2E whispered speech ASR and understand what acoustic characteristics of whispered speech have the largest effect on whispered speech ASR performance. Our proposed signal processing-based normal-to-whisper conversion method was used to create pseudo-whispered speech from three databases. Adding pseudo-whisper led to a relative WER improvement of 18\% for whispered speech and only a small WER gap with normal speech. Performance was best for the US English accent group. Comparing our results to the state-of-the-art on wTIMIT shows that our WER on whispered speech is higher than in \cite{gudepu2020whisper}; however, \cite{gudepu2020whisper} does not report which accent group from wTIMIT they use in their evaluation. Assuming they only used the US English part of wTIMIT, considering that they used large amounts of US English data from LibriSpeech for training, our results are very close to theirs (29.4\% vs. our 30.7\%) on the US English whispered speech, but using far less data. Comparing our results to those of Chang \textit{et al.} \cite{chang9383595}: both their approach and ours showed relative improvements (their 44.4\% in CER; our 18.2\% in WER); however they only report phone and character error rates, making a direct comparison not feasible in the present work.

Our final experiment shows that the lack of glottal information in whispered speech has the largest impact on whispered speech recognition.

\newpage
\bibliographystyle{IEEEbib}
\bibliography{strings,refs}

\end{document}